\documentclass[%
 reprint,
nofootinbib,
 amsmath,amssymb,
prl,
 twocolumn
]{revtex4-1}

\usepackage{graphicx}
\usepackage{bm}
\usepackage[utf8]{inputenc}
\usepackage{physics}
\usepackage[colorlinks=true,linktoc=all]{hyperref}
\usepackage{xcolor}
\usepackage{changebar}
\usepackage{mathtools}
\usepackage{mathrsfs}
\usepackage{slashed}

\DeclareMathOperator{\diag}{diag}

\begin{document}

\title{TDiff-fuelled cosmic magnetic fields}

\author{Antonio L. Maroto}
\email{maroto@ucm.es}
\affiliation{Departamento de F\'{\i}sica Te\'orica and Instituto de F\'isica de Part\'iculas y del Cosmos (IPARCOS-UCM), Universidad Complutense de Madrid, 28040 Madrid, Spain}

\author{Alfredo D. Miravet}
 \email{alfrdelg@ucm.es}
\affiliation{Departamento de F\'{\i}sica Te\'orica and Instituto de F\'isica de Part\'iculas y del Cosmos (IPARCOS-UCM), Universidad Complutense de Madrid, 28040 Madrid, Spain}


\date{\today}

\begin{abstract}

We explore the consequences of the breaking of diffeomorphism (Diff) invariance in the electromagnetic sector. We consider the breaking of Diff symmetry down to the subgroup of transverse diffeomorphisms (TDiff) and analyse its impact on the generation and evolution of cosmic magnetic fields. We show that Diff breaking induces a breaking of conformal invariance that modifies the way in which magnetic fields evolve on super-Hubble scales.  The effects of the highly conductive plasma in the evolution are also analysed. We obtain the magnetic power spectrum today and discuss the parameter regions that yield intergalactic magnetic fields compatible with current observations.

\end{abstract}

\maketitle


Magnetic fields are everywhere in the Universe, yet their origin remains a mystery to this day \cite{Durrer:2013pga}. The most agreed-upon hypothesis is that $\mu$G magnetic fields in galaxies and clusters originated via dynamo amplification \cite{Parker:1955zz} of a primordial seed, although the origin of this seed or the details of the amplification mechanism also belong to the unknown. The observation of intergalactic magnetic fields (IGMFs), which inhabit the voids of the large-scale structure and thus have not been able to undergo any sort of dynamo amplification, is key in understanding all the elements in magnetogenesis. 

The strength of large-scale IGMFs is constrained to a certain range, with almost scale-invariant bounds coming from two different types of observations: On the one hand, TeV gamma rays originated in blazars (observed at low redshift $z\sim 0.2$) create electron-positron pairs that cascade into less energetic secondary gamma rays. The absence of secondary gamma rays in the GeV range contributing to the source flux is believed to be due to the presence of magnetic fields, which bend the trajectory of charged particles. In \cite{Neronov:2010gir}, using data from Fermi/LAT telescope, a lower bound on intergalactic magnetic fields of $B \geq 10^{-16}$ G was obtained, which is independent of correlation length for sufficiently large scales $\lambda_B \geq 1$ Mpc. When considering cascade suppression due to the time delay of the signal, a more conservative lower bound is found $B \geq 10^{-18}$ G \cite{Dermer:2010mm}. On the other hand, observations of large-scale anisotropies in the CMB can be translated into an upper limit for the magnetic field strength \cite{Barrow:1997mj}, which depends on the tilt of the primordial magnetic field power spectrum \cite{Paoletti:2010rx}. 

One of the most interesting avenues for the generation of primordial seeds is the excitation of electromagnetic vacuum quantum fluctuations during inflation. However, as shown in \cite{Turner:1987bw}, the conformal triviality of Maxwell electromagnetism in Robertson-Walker backgrounds forces the 
magnetic field energy density to decay as $\rho_B \propto a^{-4}$ irrespective of the wavelength or plasma effects. Breaking conformal invariance thus becomes mandatory for any 
successful solution \cite{Grasso:2000wj, Demozzi:2009fu}. In this respect, in recent years,  mainly motivated by the success of unimodular gravity \cite{Einstein:1919gv, Unruh:1988in, Henneaux:1989zc, Kuchar:1991xd, Alvarez:2005iy, Jirousek:2018ago, Carballo-Rubio:2022ofy} as a possible solution to the vacuum energy problem \cite{Ellis:2010uc}, the interest in gravitational theories that break diffeomorphism (Diff) invariance has grown \cite{Alvarez:2006uu,Pirogov:2011iq, Bello-Morales:2023btf,Bello-Morales:2024vqk}. In particular, the breaking of Diff symmetry down to transverse diffeomorphisms (TDiff) in the gravitational couplings of matter fields has been explored in detail in the case of scalar fields  \cite{Maroto:2023toq, Jaramillo-Garrido:2023cor, Alonso-Lopez:2023hkx,Jaramillo-Garrido:2024tdv} and for abelian gauge fields  \cite{Maroto:2024mkx}. There it was found that, while the small-scale (sub-Hubble) phenomenology remains unchanged, modes beyond the Hubble horizon can evolve differently than in Diff theories.
In addition, for gauge fields, breaking down to TDiff also induces conformal invariance breaking. Precisely, this work aims to analyse the potential consequences of breaking Diff symmetry on the generation and evolution of cosmic magnetic fields.


Our starting point is the Abelian gauge field action \cite{Maroto:2024mkx}
\begin{equation}
    S[A_\mu] = - \frac{1}{4} \int \dd[4]{x} f(g)   F_{\mu\nu} F^{\mu\nu},
\end{equation}
which is invariant under transverse diffeomorphisms $\tilde x^\mu = x^\mu + \xi^\mu$, $\partial_\mu \xi^\mu = 0$, a subgroup of full diffeomorphisms comprised of volume-preserving transformations. 
Here, $F_{\mu\nu} = \partial_\mu A_\nu - \partial_\nu A_\mu$ is the field strength tensor and $f(g)$ is an arbitrary function of the metric determinant $g = |\det g_{\mu\nu}|$. Notice that for $f(g)=\sqrt{g}$ the standard Diff-invariant theory is recovered.

The gauge field follows the equations of motion
\begin{equation}\label{eq:general_eom}
  \nabla_\mu\left[ \frac{f(g)}{\sqrt{g}} F^{\mu\nu} \right] = 0. 
\end{equation}
which for the magnetic field $\vb{B}$ with components $B_i = \frac{1}{2}\epsilon_{ijk} F_{jk}$ in a Robertson-Walker background\footnote{The time coordinate $\tau$ cannot be set a priori to cosmological or conformal time, as time reparametrisations are not TDiff.}
\begin{equation}\label{eq:RWab}
    \dd{s}^2 = b(\tau)^2 \dd{\tau}^2 - a(\tau)^2 \dd{\vb{x}}^2,
\end{equation}
implies the following equation of motion 
\begin{equation}\label{eq:eom_B}
    \vb{B}'' - \frac{b^2}{a^2}\nabla^2\vb{B} + \left[ (6f_1 - 2)\frac{a'}{a}+(2f_1-2)\frac{b'}{b}\right] \vb{B}' = 0,
\end{equation}
where $f_1 = \frac{\dd\log f}{\dd \log g}$ ($f_1 \to 1/2$ to recover the Diff case) and primes indicate derivatives with respect to $\tau$. The electric field $\vb{E}$ with components $E_i = F_{0i}$ can be derived from Faraday's law
\begin{equation}\label{eq:faraday}
    \vb{B}' + \nabla\cross\vb{E} = 0.
\end{equation}

In this model, the energy density of the electromagnetic field as given by the stress-energy tensor is
\begin{equation}\label{eq:rho}
    \rho = \frac{f}{\sqrt{g}}\left(\frac{1-f_1}{a^2 b^2} \vb{E}^2 + \frac{f_1}{a^4} \vb{B}^2\right),
\end{equation}
which indicates that the electric and magnetic fields gravitate differently for $f_1\neq 1/2$.

In the super-Hubble regime, a solution for the electric and magnetic fields can be readily obtained for a power law $f(g) = g^\alpha$, with $\alpha$ a constant parametrising the Diff breaking. The most general solution that follows the equations of motion is $\vb{E} = \vb{E_0} a^{2-6\alpha}(\tau) b^{2-2\alpha}(\tau)$ for the electric field, with $\vb{E_0}$ a constant vector, and a constant magnetic field $\vb{B}$.
By inserting these solutions into the energy density \eqref{eq:rho}, we see that the electric and magnetic energy densities scale differently, with the electric part scaling as
\begin{equation}\label{eq:rhoE_propto}
    \rho_E \propto b^{1-2\alpha} a^{-1-6\alpha}
\end{equation}
and the magnetic one as 
\begin{equation}\label{eq:rhoB_propto}
    \rho_B \propto b^{2\alpha-1} a^{6\alpha - 7}.
\end{equation}
Note that in the Diff-invariant case $\alpha=1/2$, the dependence on $b$ is erased and both scale as $\rho \propto a^{-4}$ as expected. Therefore, for $\alpha\neq 1/2$, it is possible to have a magnetic field that scales differently to the conformally invariant scenario.


In what follows, we obtain the power spectrum of the magnetic field that arises as quantum fluctuations during inflation. Following \cite{Turner:1987bw}, we evaluate the spectrum in the Bunch-Davis vacuum for modes well inside the Hubble radius and then match it for super-Hubble modes at horizon crossing. 

In the covariant quantisation approach (see \cite{Maroto:2024mkx} for more details),  the gauge field can be decomposed as follows\color{black}
\begin{equation}\label{eq:vf_modes}
    A_\mu(x) = \int \frac{\dd[3]{\vb{k}}}{(2\pi)^{3/2}} \sum_\lambda \left[ a_{\vb{k}\lambda} A_{\mu,k\lambda}(x) + a_{\vb{k}\lambda}^\dagger A_{\mu,k\lambda}^*(x) \right],
\end{equation}
where the sum spans the four polarisations $\lambda=0,\dots, 3$, two of which are unphysical. The usual canonical commutation relations imply
\begin{equation}
    [a_{\vb{k}\lambda}, a^\dagger_{\vb{k}'\lambda'}] = -\eta_{\lambda \lambda'} \delta^{(3)}(\vb{k}-\vb{k'}),
\end{equation}
with $\eta_{\mu\nu} = \diag(+,-,-,-)$ and all other creation and annihilation operator commutation relations zero.

In the WKB approximation, well inside the Hubble radius, the equations of motion are solved by the following expression for the positive-frequency modes \cite{Maroto:2024mkx}
\begin{equation}\label{eq:A_wkb_normalised}
    A_{\mu, k\lambda}(x) = U_{\mu,k\lambda} e^{i\theta_{\vb{k}\lambda}} = \sqrt{\frac{b^2}{2f\omega_k}} u_{\mu,k\lambda} e^{i\vb{k}\cdot \vb{x} - i\int^\tau \omega_k(\tau')\dd{\tau'} },
\end{equation}
where the polarisation vectors satisfy $u_{\mu,k\lambda} k^\mu = 0$ and $u^*_{\mu,k\lambda} u^\mu_{k\lambda} = \eta_{\lambda\lambda}$ with the usual dispersion relation
\begin{equation}\label{eq:dispersion_relation}
    \omega_k^2 = \frac{b^2}{a^2} \vb{k}^2.
\end{equation} 
The Bunch-Davies vacuum $|0\rangle$ is thus defined by
\begin{equation}
	a_{\vb{k}\lambda} | 0 \rangle = 0,\quad \forall \vb{k},\lambda.
\end{equation}

Using the mode solution in \eqref{eq:A_wkb_normalised}, we compute the vacuum expectation value $\langle \rho_B \rangle$ from the magnetic part of \eqref{eq:rho}. The magnetic power spectrum $\rho_B(k)$ is defined as the energy density per log interval as usual, so that
\begin{equation}
    \langle\rho_B  \rangle = \int\frac{\dd{k}}{k} \rho_B(k),
\end{equation}
which yields for sub-Hubble modes
\begin{equation}\label{eq:rhoB_subH}
    \rho_B(k) = \frac{f_1}{2\pi^2} \frac{k^4}{a^4}.
\end{equation}

Similarly, one can obtain the following expression for the electric power spectrum
\begin{equation}\label{eq:rhoE_subH}
    \rho_E(k) = \frac{1-f_1}{2\pi^2} \frac{k^4}{a^4}.
\end{equation}

These energy densities scale as $a^{-4}$, as expected for a free massless vector field. Also, the total energy density (i.e. the sum of the electric and magnetic contributions) is independent of the TDiff coupling, as the factors $f_1$ cancel out.  This agrees with the fact that, in the sub-Hubble regime and when the geometric optics approximation is applicable, the phenomenology of a TDiff-invariant field is analogous to its Diff-invariant counterpart \cite{Maroto:2024mkx}.

At horizon crossing $k=aH$, which is the very last moment when the sub-Hubble regime applies, the power spectrum reads $\rho_B(k)\sim H_I^4$, which is nearly scale-invariant in a typical inflationary scenario. After horizon crossing, the energy density acquires the super-Hubble scaling \eqref{eq:rhoB_propto}, which must be complemented by an expression for the lapse function in terms of the scale factor $b=b(a)$. Such an expression can be derived from the energy conservation equations $\nabla_\mu T^{\mu\nu}=0$, which are not fulfilled on solutions to the field equations but must be for Einstein's equations as per Bianchi identities. For sub-Hubble modes, the energy-momentum tensor is conserved for any $b$ \cite{Maroto:2024mkx}, so we need to look at super-Hubble modes only. For these, the ratio of electric to magnetic energy density is given by
\begin{equation}
    \left. \frac{\rho_E(k)}{\rho_B(k)} \right|_\mathrm{super-Hubble} = \frac{1-\alpha}{\alpha} \left( \frac{g}{g_k} \right)^{1-2\alpha},
\end{equation}
with $g_k = g(a_k, b_k)$ the metric determinant at horizon crossing. Here we can see that, depending on the value of $\alpha$ (for $\alpha\neq 1/2$), either the electric or magnetic energy density will dominate not long after the mode crosses the horizon. The conditions for magnetic field domination, while ensuring a positive magnetic energy density, are:
\begin{align*}
    |\rho_B| \gg |\rho_E| : \left\{\begin{array}{ll}
        g'>0, &  \alpha>1/2 \\
        g'<0, &  0<\alpha<1/2 
    \end{array}\right.
\end{align*}

In this case, the energy conservation yields \cite{Maroto:2024mkx}
\begin{equation}
    C_Ba^4 = g^\alpha,
\end{equation}
for $\alpha \neq 1/2$ with $C_B$ a constant. This implies $b \propto a^{(2-3\alpha)/\alpha}$, so the metric determinant $g = b^2a^6 \propto a^{4/\alpha}$, which satisfies $g'>0$ if $a'>0$ (expanding universe) and $\alpha>0$. Therefore, considering the magnetic condition only is correct for expanding universes and $\alpha>1/2$. In this case, the magnetic energy density scales as 
\begin{align}
\rho_B \propto a^{-2/\alpha} \label{eq:rhoB_propto2}
\end{align}
whereas the electric energy  density \eqref{eq:rhoE_subH} satisfies
$\rho_E \propto a^{2/\alpha-8}$. Thus we see that for $\alpha>1/2$ the magnetic energy density on super-Hubble scales  dilutes more slowly than in the Diff invariant case.


Up to this point, we have only considered the free electromagnetic field in an expanding universe. However, for most of the known thermal history of the Universe, there has been a large density of electrically charged particles, which results in a Universe featuring a large conductivity $\sigma_c$. In order to account for this fact, we need to introduce the interactions of the gauge field.

Let us consider the gauge coupling to a spinor field, which we write in the following way
\begin{multline}
    S = \int \dd[4]{x} \left( f(g) \left[ -\frac{1}{4} F_{\mu\nu} F^{\mu\nu} + \frac{i}{2}(\bar\Psi \slashed{\mathfrak{D}} \Psi - \slashed{\mathfrak{D}} \bar\Psi \Psi) \right] \right.\\
    \left. + f_C(g) \left[-q \bar\Psi \slashed{A} \Psi\right]\right).
\end{multline}
where $\mathfrak{D}_\mu$ denotes the covariant derivative acting on spinor fields and we have allowed for a different coupling function $f_C(g)$ in the interaction term. 
In order to study the coupling between the two fields, we first need to write the free action in its canonical form. Let us write the free action as
\begin{equation}
    S = \int \dd[4]{x} f(g) \mathcal{L}[\Phi, g_{\mu\nu}] = \int \dd[4]{x} \sqrt{g} \frac{f(g)}{\sqrt{g}}  \mathcal{L}[\Phi, g_{\mu\nu}],
\end{equation}
which depends on the metric tensor and the fields $\Phi=\{ A_\mu, \Psi  \}$. Let us consider the transformation to a system of coordinates $\tilde x^\mu$ in which the metric becomes the flat metric $\tilde{g}_{\mu\nu}=\eta_{\mu\nu}$ in a Minkowskian neighbourhood $\mathcal{M}$. In this new system of coordinates, since $\dd[4]{x}\sqrt{g}$ and $\mathcal{L}$ are both scalars under general coordinate transformations, the action reads
\begin{equation}
    S = \int_{\mathcal{M}} \dd[4]{\tilde{x}} \frac{f(g)}{\sqrt{g}} \mathcal{L}[\tilde \Phi, \eta_{\mu\nu}],
\end{equation}
where the transformation of the factor $f(g)/\sqrt{g}$ is unknown unless we specify an explicit form for $f(g)$, which will determine its scalar weight. Therefore, we leave this factor as is, which should be rewritten in terms of the new coordinates. Since the free action is quadratic in the fields, the canonically normalised fields read
\begin{equation}
    \hat \Phi = \left( \frac{f(g)}{\sqrt{g}} \right)^{1/2} \Phi.
\end{equation}

 In addition, in order to prevent the evolution of the coupling constant, which could allow it into the strong coupling regime\footnote{Notice that for $\alpha>1/2$, as required for magnetic amplification, $f(g)/\sqrt{g}$ grows as $a^{(4-2/\alpha)}$ so that if today $g_0=1$ we would have a strong coupling problem during inflation \cite{Demozzi:2009fu}}, the coupling function of the interaction term needs to be
\begin{equation}
    f_C(g) = \frac{f^{3/2}(g)}{g^{1/4}},
\end{equation}
which also prevents violation of local position invariance for the charge from happening. With this choice, the action  written in terms of the canonically normalised  fields finally reads
in the Minkowskian neighbourhood 
\begin{equation}\label{eq:action_with_hats}
    S = \int_{\mathcal{M}} \dd[4]{x} \left( -\frac{1}{4} \hat F_{\mu\nu} \hat{F}^{\mu\nu} + \frac{i}{2}(\bar{\hat\Psi} \slashed{\partial} \hat\Psi - \slashed{\partial} \bar{\hat\Psi} \hat\Psi) - q\bar{\hat\Psi} \hat{\slashed{A}} \hat\Psi \right),
\end{equation}
which is gauge invariant in $\mathcal{M}$, i.e. at the leading adiabatic order in which terms
involving metric derivatives are negligible compared to
those involving derivatives of the fields.

Conductivity is accounted for by introducing a current proportional to the electric field as per Ohm's law \cite{Tsagas:2004kv}
\begin{equation}
	\hat{j}^\mu - u^\mu u_\nu \hat{j}^\nu = \sigma_c \hat{F}^{\mu\nu} u_\nu,
\end{equation}
where $\sigma_c$ is the conductivity of the plasma. Naturally, the quantities that appear in Ohm's law are the canonical fields, and the canonical  current $\hat{j}^\mu$ can be read from the action \eqref{eq:action_with_hats}
\begin{equation}
    \hat{j}^\mu = q \bar{\hat\Psi} \gamma^\mu \hat\Psi.
\end{equation}

In terms of the original fields, Ohm's law reduces to $j^i = b \sigma_c F^{i0}$ for a neutral plasma $u_\mu j^\mu = 0$, which introduces a conductivity term in the right-hand side of \eqref{eq:eom_B}
\begin{equation}
    \vb{B}'' - \frac{b^2}{a^2}\nabla^2\vb{B} + \left[ (6f_1 - 2)\frac{a'}{a}+(2f_1-2)\frac{b'}{b}\right] \vb{B}' = -b \sigma_c \vb{B}'.
\end{equation}

A very high conductivity $\sigma_c\to\infty$ requires the magnetic field to be constant $\vb{B}'\to 0$, which makes the magnetic field behave as in the super-Hubble regime. This also applies in the sub-Hubble regime as long as $a \sigma_c \gg k$, which can be identified as the overdamped regime.

Following \cite{Turner:1987bw}, let us study the effect of conductivity on the different modes. We are interested in magnetic fields of comoving size $\lambda = 2\pi/ k \gtrsim 0.1$ Mpc, so we shall estimate whether conductivity dominates before the present moment for such modes.  Starting at some point during reheating, conductivity is very high until electron-positron annihilation at $T_\mathrm{ann}= m_e \simeq 0.5 \,\mathrm{MeV}$. Conductivity can be estimated as $\sigma_c \sim X_e m_e / e^2$, where $X_e$ is the ratio of free electrons per photon and the charge of the electron is $e^2 = 4\pi \alpha \sim 1/10$. The lowest value $X_e$ acquires is $\mathcal{O}(10^{-13})$, which happens after recombination, so $\sigma_c \gtrsim 10^{-12} m_e$. We compare this value to the  wavenumber of modes that enter the horizon up to today
\begin{equation}
    \left.\frac{\sigma_c}{k/a}\right|_\mathrm{sub} \gtrsim \frac{\sigma_c}{(k/a)_\mathrm{ann}} = \frac{\sigma_c}{H_\mathrm{ann}} \sim \frac{\sigma_c}{T_\mathrm{ann}^2/M_P} \gtrsim 10^{11} \gg 1,
\end{equation}
where ``sub'' refers to sub-Hubble modes only, so conductivity stays high until today.


\begin{figure}
    \centering
    \includegraphics[width=\linewidth]{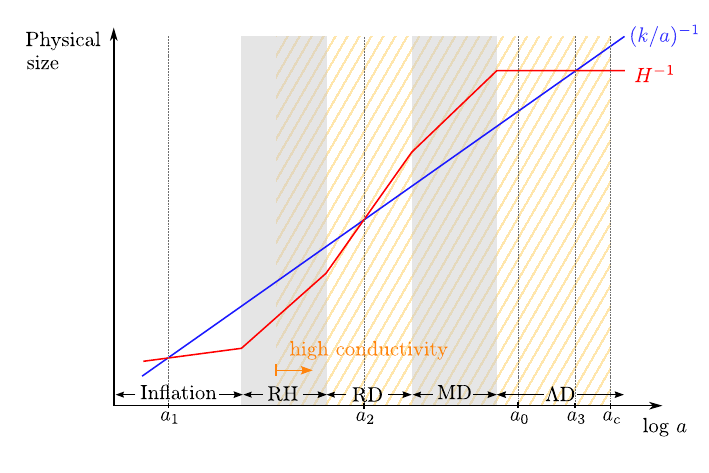}
    \caption{Evolution of the Hubble radius (in red) and the physical wavelength of a mode with comoving wavenumber $k$ (in blue). The yellow striped region corresponds to a high conductivity period, which starts during reheating and ends at scale factor $a_c$. Scale factors for $n$-th horizon crossing $a_n$ and today $a_0$ are also labelled.}
    \label{fig:scales_plot}
\end{figure}

Following Fig. \ref{fig:scales_plot}, let us review the complete evolution of a mode $\rho_B(k)$. During inflation, the fluctuations of the magnetic field get excited and the power spectrum acquires the following value at the first horizon crossing $a=a_1=k/H_I$:
\begin{equation}
    \left.\rho_B(k)\right|_{a=a_1} = \rho_B (k=aH_I) = \frac{f_1}{2\pi^2} H_I^4.
\end{equation}

Thus, if the magnetic field dominates over the electric field, this energy density evolves as \eqref{eq:rhoB_propto2} $\rho_B\propto a^{-2/\alpha}$, with $\alpha>1/2$.  This scaling holds until today for intergalactic-size modes, either because the mode is super-Hubble or because the conductivity is very high so we just need to evaluate it at the moment $a$ we are interested in. Scaling the energy density, we have
\begin{equation}
    \rho_B(k) = \frac{\alpha}{2\pi^2} H_I^4 \left( \frac{k}{a H_I}\right)^{2/\alpha},
\end{equation}
with the particular case $a=a_0=1$ for the energy density today. 

The magnetic power spectrum can also be defined as \cite{Durrer:2013pga}
\begin{equation}
    P_B(k) = \frac{2\pi^2}{k^3} \rho_B(k) \propto k^{\frac{2}{\alpha}-3},
\end{equation}
which displays a tilt
\begin{equation}\label{eq:nb_alpha_tilt}
    n_B = \frac{2}{\alpha} - 3.
\end{equation}

The magnetic field intensity today at a certain comoving scale $\lambda$, which follows from \eqref{eq:rho}, reads
\begin{equation}\label{eq:b_lambda}
    B_\lambda = \left. \sqrt{\frac{\rho_{B,0}(k)}{\alpha}} \right|_{k=\frac{2\pi}{\lambda}}.
\end{equation}

\begin{figure}
    \centering
    \includegraphics[width=\linewidth]{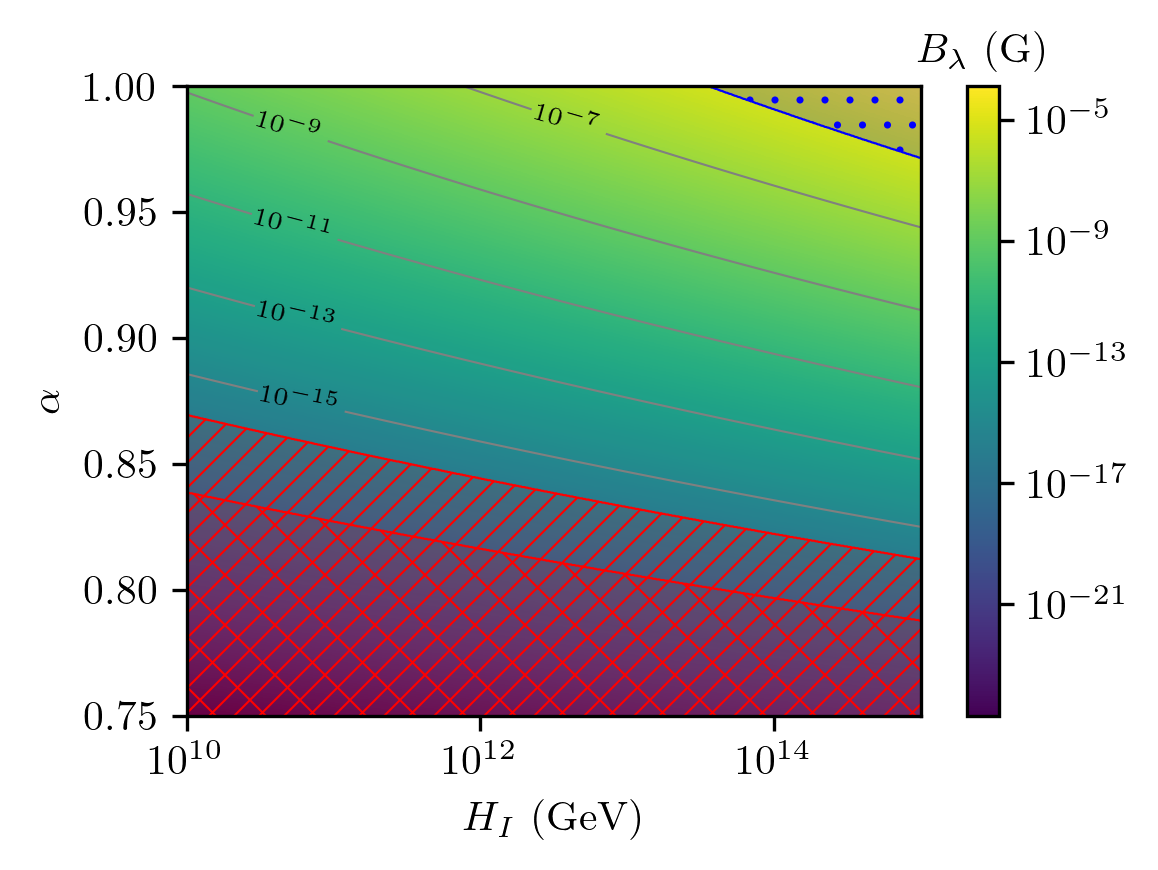}
    \caption{Magnetic field intensity $B_\lambda$ today (in Gauss) for a mode of wavelength $\lambda = 1\mathrm{ Mpc}$, as a function of the inflation scale $H_I$ and the Diff-breaking paramater $\alpha$. The upper right blue dotted region indicates the region excluded by large-scale CMB observations. The bottom red striped region corresponds to values $B_{1\mathrm{ Mpc}} \lesssim 10^{-16}$ and  $B_{1\mathrm{ Mpc}} \lesssim 10^{-18}$ G, excluded by blazar observations, depending on whether the least conservative bound is chosen or not, respectively (see text for details). We do not plot all values down to $\alpha = 1/2$ as they are excluded by blazar observations.}
    \label{fig:b_parameters}
\end{figure}

By substituting the energy density, one can obtain the following expression
\begin{equation}
	B_\lambda = \frac{1.3\mu\text{G}}{(2.5\times 10^{50})^{(1-\alpha)/\alpha}} \left(\frac{H_I}{10^{13}\mathrm{GeV}}\right)^\frac{2\alpha-1}{\alpha} \left(\frac{\mathrm{Mpc}}{\lambda}\right)^{1/\alpha},
\end{equation}
which allows for values of the magnetic field of the order of the $\mu$G at galactic scales for $\alpha \simeq 1$.
These expressions encapsulate the following features of the magnetic power spectrum:
\begin{enumerate}
    \item The spectrum can be either red or blue-tilted, with $-1 \leq n_B \leq 1$ for the range $\frac{1}{2} \leq \alpha \leq 1$. 
    \item The energy density depends on the inflation scale as $\rho_B \propto H_I^{4-2/\alpha}$, so it grows larger for large inflation scales. On top of that, the larger $\alpha$ is, the more it grows with the inflation scale. This is depicted in Fig. \ref{fig:b_parameters}.
    \item The magnetic field intensity is enhanced for larger $\alpha$, which is caused by the ratio $k/H_I$ being typically very small. This can also be seen in Fig. \ref{fig:b_parameters}.
\end{enumerate}

In Fig. \ref{fig:b_parameters}, we plot the predicted magnetic field intensity today \eqref{eq:b_lambda} for a scale of $\lambda = 1\mathrm{Mpc}$ together with the current large-scale bounds on IGMFs. We see that a TDiff-invariant electromagnetic sector could produce the observed values for IGMFs with a minimal inflationary scenario, provided the Diff-breaking parameter $\alpha\gtrsim 0.8$ and the inflation scale is sufficiently high.

The lower bound in Fig. \ref{fig:b_parameters} is given by blazar observations by Fermi/LAT \cite{Neronov:2010gir}, which are low-redshift events, so we can use the bounds directly in our model. The upper bounds, however, come from limits on CMB anisotropies \cite{Barrow:1997mj} which have been scaled up to today $\rho_B^\mathrm{CMB} = \rho_{B,0} a_\mathrm{CMB}^{-4}$. Since the magnetic energy density scales differently in our model $\rho_B^\mathrm{CMB} = \rho_{B,0} a_\mathrm{CMB}^{-2/\alpha}$, we have adapted the bound in order to reflect the different dilution between the last scattering surface and the present time. Moreover, these CMB bounds depend on magnetic power spectrum tilt, which is related to our parameter $\alpha$ \eqref{eq:nb_alpha_tilt}. In \cite{Paoletti:2010rx}, the dependence of the CMB bounds on the tilt was examined, which showed that blue-tilted power spectra are more restricted. In particular, they found $B_{1\mathrm{Mpc}}\lesssim 1$ nG for $n_B=1$ and $B_{1\mathrm{Mpc}}\lesssim 6$ nG for $n_B=-1$ at $2\sigma$ level, with a somewhat linear regression between the two. We have translated these limits when computing the CMB limits in the plot. Thus we see that the maximum value that TDiff models can produce in the $1/2 < \alpha \leq 1$ range and that is compatible with the CMB bound is $B_{1\mathrm{Mpc}} = 2.6\mu\text{G}$, which is of the order of observed galactic magnetic fields.

In conclusion, the breaking of Diff invariance offers an interesting framework to analyse the problem of the origin of cosmic magnetic fields. The estimates presented in this work suggest that a restricted (TDiff) symmetry of the electromagnetic sector could play a role in their evolution and possible amplification after inflation.

\acknowledgments{This work has been supported by the MICIN (Spain) Project No. PID2022-138263NB-I00 funded by MICIU/AEI/10.13039/501100011033 and by ERDF/EU. A. D. M. acknowledges financial support by the MICIU (Spain) through a Formación de Profesorado Universitario (FPU) fellowship FPU18/04599}

\bibliography{tdiff-vec.bib}

\end{document}